\documentclass{article}
\usepackage{ijcai24}

\usepackage{times}
\usepackage{soul}
\usepackage{url}
\usepackage[hidelinks]{hyperref}
\usepackage[utf8]{inputenc}
\usepackage[small]{caption}
\usepackage{graphicx}
\usepackage{amsmath}
\usepackage{amsthm}
\usepackage{booktabs}
\usepackage{algorithm}
\usepackage{algorithmic}
\usepackage[switch]{lineno}
\usepackage{multirow}
\usepackage{amsfonts}
\usepackage{enumitem}
\usepackage{mathrsfs}
\usepackage{tikz}
\usepackage{pgfplots}
\usepackage{pifont}

\title{ReliaAvatar: A Robust Real-Time Avatar Animator \\with Integrated Motion Prediction}

\author{
    Bo Qian, Zhenhuan Wei, Jiashuo Li, Xing Wei\thanks{Corresponding author: Xing Wei.}
    \affiliations
    School of Software Engineering, Xi'an Jiaotong University
    \emails
    \{qb990531, zh-wei, xjtuljs\}@stu.xjtu.edu.cn,weixing@mail.xjtu.edu.cn
}

\begin{document}

\maketitle

\begin{abstract}
Efficiently estimating the full-body pose with minimal wearable devices presents a worthwhile research direction. 
Despite significant advancements in this field, most current research neglects to explore full-body avatar estimation under low-quality signal conditions, which is prevalent in practical usage.
To bridge this gap, we summarize three scenarios that may be encountered in real-world applications: standard scenario, instantaneous data-loss scenario, and prolonged data-loss scenario, and propose a new evaluation benchmark.
The solution we propose to address data-loss scenarios is integrating the full-body avatar pose estimation problem with motion prediction. Specifically, we present \textit{ReliaAvatar}, a real-time, \textbf{relia}ble \textbf{avatar} animator equipped with predictive modeling capabilities employing a dual-path architecture.
ReliaAvatar operates effectively, with an impressive performance rate of 109 frames per second (fps).
Extensive comparative evaluations on widely recognized benchmark datasets demonstrate Relia\-Avatar's superior performance in both standard and low data-quality conditions.
The code is available at \url{https://github.com/MIV-XJTU/ReliaAvatar}.
\end{abstract}

\section{Introduction}
Virtual reality, augmented reality (AR), and mixed reality (MR) are rapidly evolving fields that offer new dimensions for human communication and interaction. A critical aspect of enhancing these immersive experiences lies in the ability of models to generate seamless and realistic avatar motions, ideally using user-friendly devices such as head-mounted displays (HMD) or non-wearable WiFi devices~\cite{person3dyan}. Despite notable advancements in using sparse observations to animate full-body avatars, a significant research gap exists to effectively drive avatars in scenarios plagued by low-quality signals.

\begin{figure}[t]
    \centering
    \includegraphics[width=\linewidth]{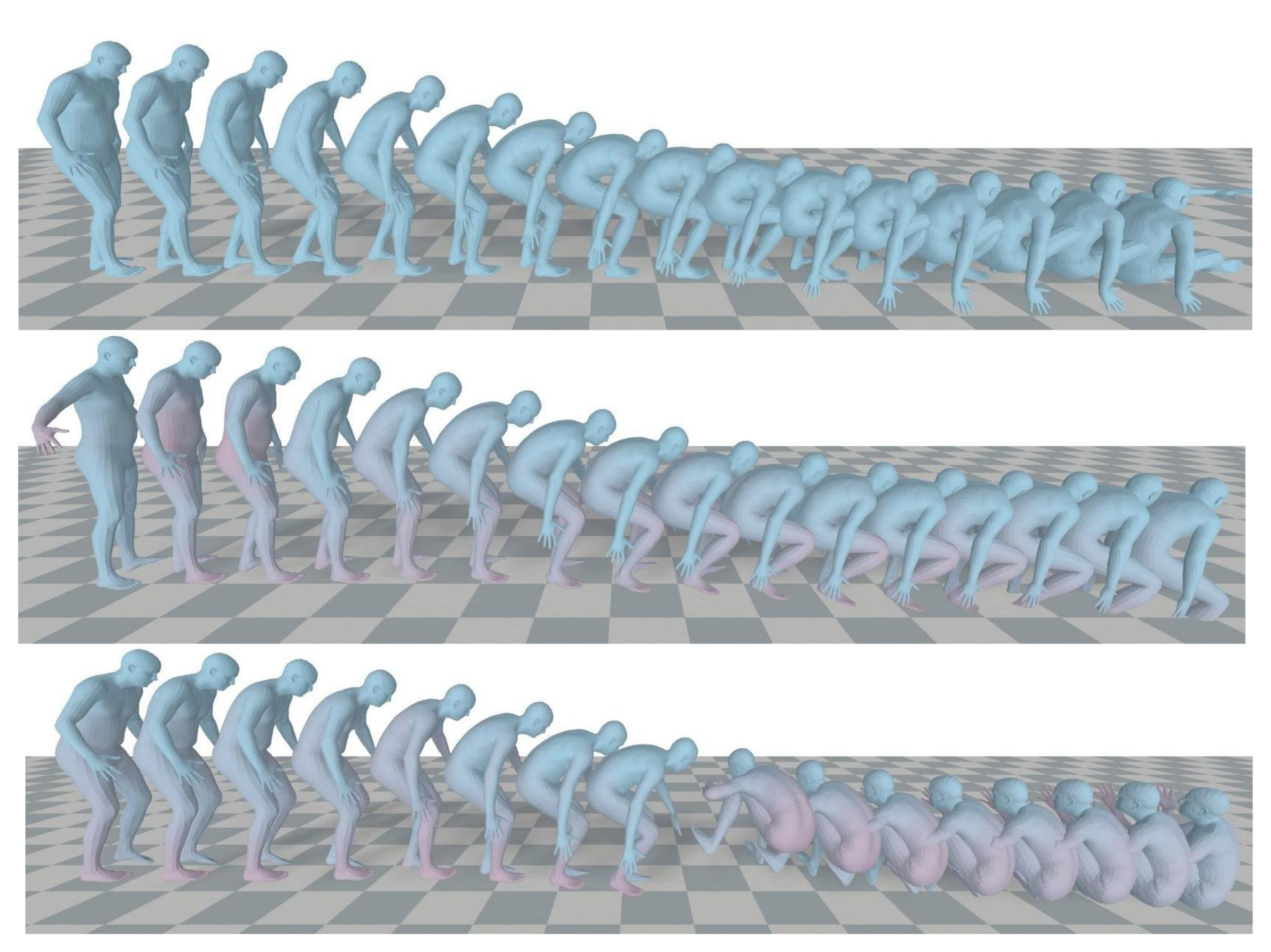}
    \caption{Visualization in the context of prolonged data-loss scenario. We mask out the latter half of a sample, which consisted of 80 frames and depicted ``crouching". The first row represents the ground truth, the second row represents the response of ReliaAvatar, and the third row represents the response of AvatarPoser. The visualization clearly indicates that ReliaAvatar can operate effectively in prolonged data loss scenarios with only minor distortions. In contrast, AvatarPoser completely fails to perform in this scenario.}
    \label{fig:vishead}
    \vspace{-12pt}
\end{figure}

In practical applications, poor-quality signals are a common challenge, particularly when wearable devices are intended to be as convenient as possible. 
Various factors, including network fluctuations, occlusion in motion capture systems, and limited visibility of interactive handles in HMDs, can degrade signal integrity. Previous works such as AGRoL~\cite{du2023avatars} and HMD-NeMo~\cite{aliakbarian2023hmd} have addressed some aspects of these challenges. However, systematic exploration and comprehensive solutions for diverse data-loss scenarios remain underdeveloped.

\begin{table}[t]
    \centering
    \begin{tabular}{c|ccc}
    \toprule
    Method & Standard &
    Instantaneous& Prolonged  \\
    \midrule
    Other Methods &  \checkmark & \ding{53} &\ding{53}\\
    ReliaAvatar & \checkmark & \checkmark & \checkmark \\
    \bottomrule
    \end{tabular}
    \caption{Ability to handle different scenarios.}
    \label{tab:diff}
    \vspace{-15pt}
\end{table}
In this work, we delve into potential data-loss scenarios that could occur in real-world applications. We specifically identify and focus on two scenarios: instantaneous data-loss and prolonged data-loss scenarios. Our response to these challenges is two-fold, encompassing both model architecture and training methodology innovations. We introduce \textbf{\textit{Relia\-Avatar}}, a real-time, robust, autoregressive Avatar Animator, integrating full-body joint motion prediction with an innovative training approach that simulates data-loss conditions.

We develop two distinct data-loss scenarios: \textbf{instantaneous}, resembling tracker signal loss like network packet loss, and \textbf{prolonged}, simulating long-last loss of specific motion capture joints, such as a hand-held controller's prolonged invisibility. Our model operates through two pathways: a regression pathway for conventional full-body avatar pose estimation tasks and a prediction pathway that predicts motions in the absence of tracker signals, ensuring continuity in avatar movements. Both pathways leverage GRU-based models for feature extraction. The combined features are then transformed into a decoding token sequence representing 22 SMPL joints, which will be used in a Transformer encoder to model the inter-joint relationships. 
We also propose an autoregressive training pipeline encompassing three preprocessing methods aligned with standard, instantaneous, and prolonged data-loss scenarios. During training, each signal sequence is processed using one of these preprocessing methods, enabling the model to adapt to the challenges posed by different data loss scenarios.

We compare ReliaAvatar with existing methods in all scenarios on the AMASS benchmark dataset~\cite{mahmood2019amass}. The results showcase our model's state-of-the-art performance in both standard and data-loss scenarios. We also provide a comparison with AvatarPoser~\cite{jiang2022avatarposer} in Figure \ref{fig:vishead} in the prolonged data-loss scenario. These qualitative and quantitative results collectively demonstrate that we are the first method, as stated in Table \ref{tab:diff}, that has robustness to data quality.
Our model also demonstrates impressive performance during the online inference stage, achieving a remarkable 109 fps. This outstanding speed surpasses other avatar pose estimation methods, highlighting the superiority of our model in real-time applications.

\noindent Our contribution can be summarized as follows:
\begin{itemize}
\item We pioneer a comprehensive investigation into practical data-loss scenarios, identifying and focusing on two key scenarios: instantaneous and prolonged data loss.
\item We propose ReliaAvatar, a real-time, robust avatar animator with integrated full-body joint motion prediction, and an autoregressive training pipeline tailored for data-loss scenarios. This approach significantly enhances the model's adaptability and performance.
\item Our experimental results demonstrate that ReliaAvatar not only achieves top-tier performance in the standard scenario but also represents, to our knowledge, the first method adept at managing various data-loss scenarios effectively. Furthermore, ReliaAvatar outperforms other methods in terms of computational efficiency.
\end{itemize}

\section{Related Work}

\subsection{Full-body Avatar Pose Estimation}
In the past, motion capture systems required users to wear numerous devices, as seen in early implementations \cite{xsens2013full,vlasic2007practical}. However, this requirement often compromises the immersive experience, highlighting the need for more compact approaches. Consequently, the focus shifted to generating a full-body pose using sparse observations, a topic that has garnered significant attention and led to numerous advancements \cite{von2017sparse,huang2018deep,yi2021transpose,yang2021lobstr,jiang2022avatarposer,zheng2023realistic,du2023avatars,aliakbarian2023hmd}. These innovative methods allow users to wear tracking devices on only a limited number of joints, such as the head, hands, and pelvis, thus improving the overall user experience.
Among these developments, AvatarPoser \cite{jiang2022avatarposer} introduced a Transformer-based architecture combined with an inverse kinematics (IK) module, paving the way for more sophisticated avatar control techniques. AGRoL \cite{du2023avatars} utilized diffusion models to drive avatars, showing notable robustness against instantaneous data loss. Similarly, HMD-NeMo \cite{aliakbarian2023hmd} made strides in avatar control by focusing on scenarios where the hands are partially or entirely obscured.
Building on these foundational works, our research delves deeper into the realm of full-body avatar estimation under the conditions of low-quality data.

\subsection{3D Human Motion Prediction} 

A key innovation of our model is the motion prediction pathway in the architecture, which is used to provide additional cues to avatar pose estimation.
There are many motion prediction methods available for reference, ranging from nonlinear Markov models \cite{lehrmann2014efficient} and Restricted Boltzmann Machines \cite{taylor2006modeling}, to more recent developments like Graph Convolutional Networks \cite{ma2022progressively,mao2020history,mao2019learning} and Recurrent Neural Networks \cite{jain2016structural,liu2019towards,martinez2017human}.
In the prediction pathway, we use a GRU module\cite{cho2014learning} to extract features and a Transformer \cite{vaswani2017attention} module to model the inter-joint relationships.
The prediction pathway can be aggregated with the regression pathway at the joint-relation Transformer to form our reliable avatar animator.

\section{Method}

\subsection{Problem Definition} \label{sec:definition}

{\flushleft\textbf{Task.}}
The core objective of our task is to accurately predict the 3D full-body human poses $y^t$ at any given time-step $t$, using sparse joint signals $x^t$ and supplementary information $e^t$. This supplementary information, $e^t$, may include data derived from historical movements or other available motion signals at time-step $t$. Therefore, our objective is formally defined as follows.
\begin{equation}
\max\limits_{\theta} \mathbb{P}( y^t|x^t,e^t,f_{\theta})
\end{equation}
Traditionally, this task is approached as an upsampling challenge, where the goal is to learn the mapping from sparse joint signals to a complete full-body joint configuration.
In these conventional methods, inputs are configured as a series of continuous sparse signals over a time window to facilitate the extraction of temporal features, leading to the following model formulation.
\begin{equation}
\hat{y}^t=f_{\theta}(x^t, x^{t-1},..., x^{t-T})
\end{equation}
Here, the supplementary information $e^t$ is represented as $e^t=\{{x_i}\}_{i=t-T}^{t-1}$, with $T$ denoting the window size. 

Those approaches neglect the continuity of the prediction process, which has been proven to be crucial in visual tracking\cite{wei2023autoregressive,Bai_2024_CVPR}.
Unlike them, we treat the task as an integrative process of both upsampling and full-body joint motion prediction. To this end, we incorporate historical trajectory states as an additional input variable. We utilize a Gated Recurrent Unit (GRU) to extract temporal features from both the tracker signals and the historical trajectory states. This approach allows us to redefine $e^t$ as a composite of previous pose predictions and historical hidden states, i.e., $e^t=\{\hat{y}^{t-1},h_x^{t-1},h_y^{t-1}\}$. Consequently, our model's formulation is as follows:
\begin{equation}
\label{eq:full_formulation}
\hat{y}^t,h_x^t,h_y^t=f_{\theta}(x^t,\hat{y}^{t-1},h_x^{t-1},h_y^{t-1})
\end{equation}
Here, $h_x^{t},h_y^{t}$ represent the hidden states generated by the GRU, essential for capturing the complex temporal dynamics of human motion.

\begin{figure*}[!t]
    \centering
    \includegraphics[width=0.78\linewidth]{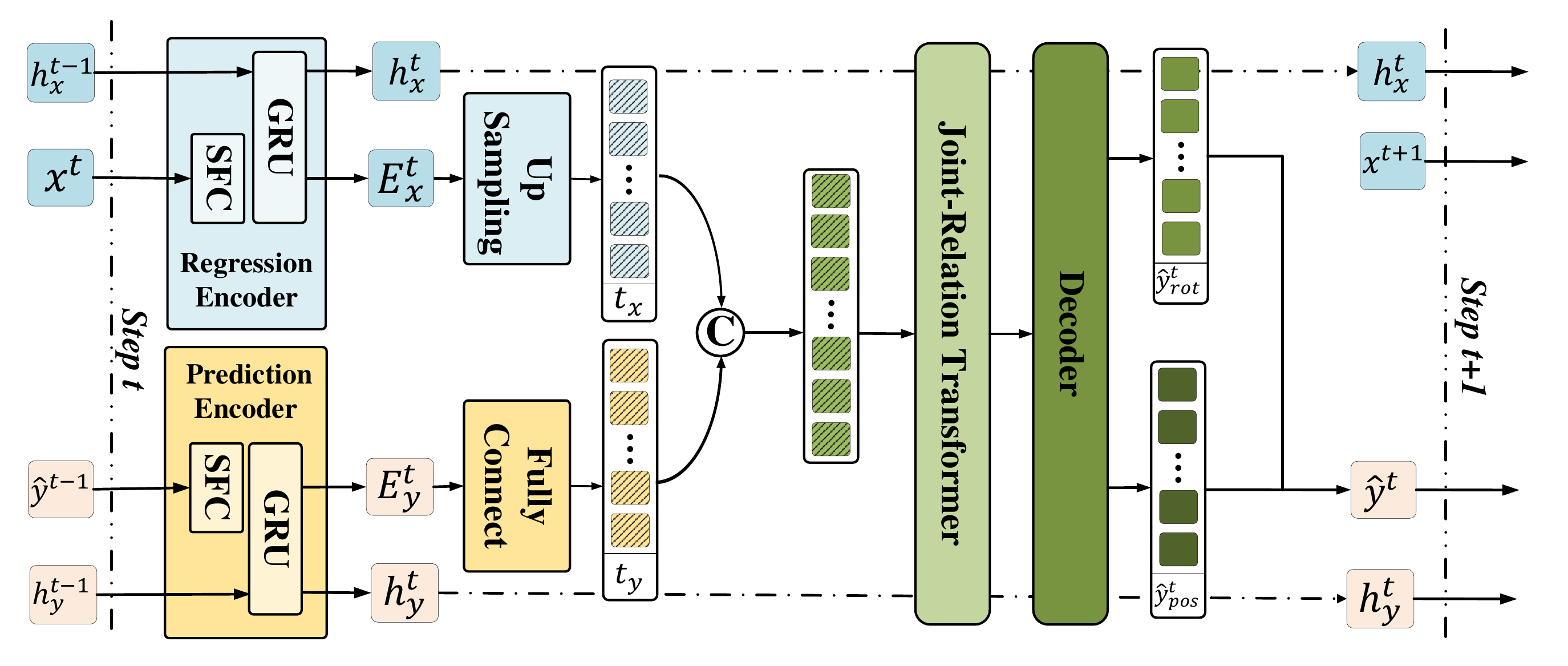}
    \caption{Illustration of our dual-pathway, autoregressive framework. ReliaAvatar has two pathways: the regression pathway (Regression Encoder$\rightarrow$Joint-Relation Transformer$\rightarrow$Decoder) and the prediction pathway(Prediction Encoder$\rightarrow$Joint-Relation Transformer$\rightarrow$Decoder). The output of ReliaAvatar at time-step $t$ forms a part of the input at time-step $t+1$. The blocks with diagonal lines as foreground (\includegraphics[height=0.9em]{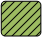},\includegraphics[height=0.9em]{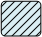},\includegraphics[height=0.9em]{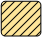}) represent tokens that signify SMPL joints, e.g., pelvis, left wrist, right ankle.}
    \label{fig:pipeline}
    \vspace{-12pt}
\end{figure*}
{\flushleft\textbf{Representation.}}
ReliaAvatar's input signals consist of four parts: the tracker signals $x^t$, hidden tracker states $h_x^{t-1}$, the historical trajectory states $\hat{y}^{t-1}$, and hidden historical states $h_y^{t-1}$. Here, $\hat{y}^{t-1}$ represents the output of the model at time-step $t-1$, and $h_x^{t-1}$ and $h_y^{t-1}$ are the hidden states returned by the GRU module.

The input tracker signals $x^t$ at time-step $t$ are composed of the orientation\footnote{The orientation is represented by the 6D representation of
rotation~\cite{zhou2019continuity}.} $x_r \in \mathbb{R}^{|\mathbb{J}|\times6}$, rotation velocity $\Delta{x}_r \in \mathbb{R}^{|\mathbb{J}|\times6}$, position $x_p \in \mathbb{R}^{|\mathbb{J}|\times3}$, and linear velocity $\Delta{x}_p \in \mathbb{R}^{|\mathbb{J}|\times3}$ of the trackers. $\mathbb{J}$ represents the set of all trackers, $\Delta{x}_p = x_p^t - x_p^{t-1}$, $\Delta{x}_r = f_{6D}(R^{-1}(x_r^{t-1})R(x_r^t))$, and $R$ and $R^{-1}$, respectively, represent converting the 6D representation of rotation into the corresponding rotation matrix and inverse matrix, where $f_{6D}$ represents converting the rotation matrix into a 6D representation. Therefore, $x^t=\{x_r^t,\Delta{x}_r^t,x_p^t,\Delta{x}_p^t\}\in \mathbb{R}^{|\mathbb{J}|\times 18}$.

The historical trajectory states $\hat{y}^{t}$ at time $t+1$ are also composed of the four parts: 22 SMPL joints' local rotation relative to their parent joints $\hat{y}_r^t \in \mathbb{R}^{22\times6}$, position $\hat{y}_p^t \in \mathbb{R}^{22\times3}$ and corresponding velocity $\Delta{\hat{y}}_r^t \in \mathbb{R}^{22\times6}$ and $\Delta{\hat{y}}_p^t \in \mathbb{R}^{22\times3}$. Therefore, the overall output can be represented as $\hat{y}^t=\{\hat{y}_r^t,\Delta{\hat{y}}_r^t,\hat{y}_p^t,\Delta{\hat{y}}_p^t\} \in \mathbb{R}^{396}$.

\begin{figure}
  \centering
  \begin{minipage}{0.23\textwidth}
    \centering
    \includegraphics[width=\linewidth]{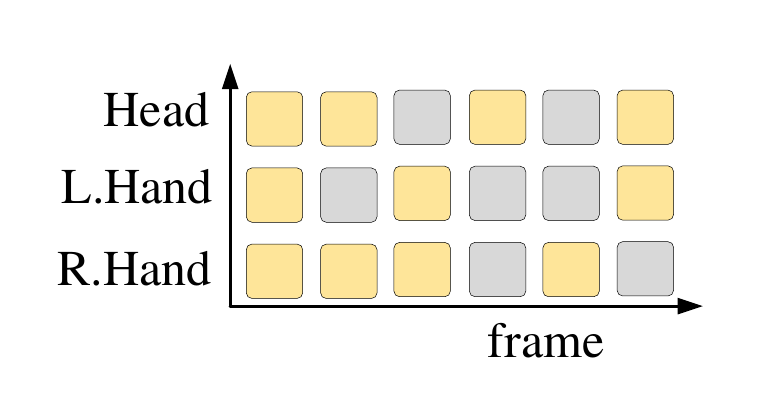}
    \vspace{-22pt}
    \caption*{\textbf{A)} Instantaneous data-loss}
    
  \end{minipage}
  \hfill
  \begin{minipage}{0.23\textwidth}
    \centering
    \includegraphics[width=\linewidth]{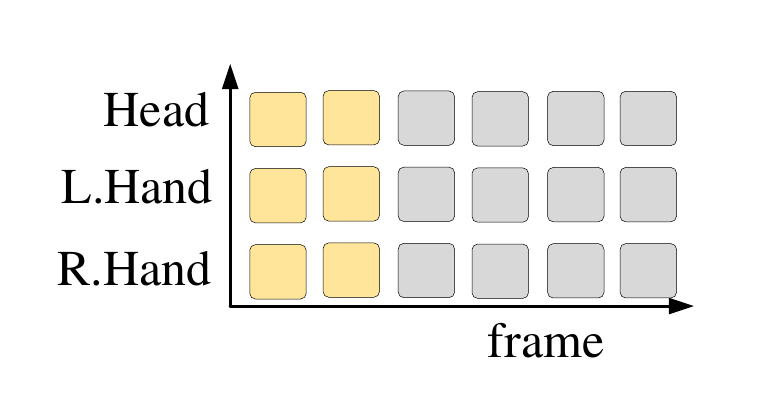}
    \vspace{-22pt}
    \caption*{\textbf{B)} Prolonged data-loss}
    
  \end{minipage}
  \vspace{-2pt}
  \caption{Illustration of data-loss scenarios.\includegraphics[height=0.95em,width=1.2em]{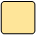} represents that the signals have been received properly, while \includegraphics[height=0.95em,width=1.2em]{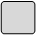} represents that the signals are lost.} 
  \label{fig:simulation}
  \vspace{-12pt}
\end{figure}

{\flushleft\textbf{Scenarios.}} 
Current models often operate under the assumption that tracker signals are consistently available, promptly received, and accurate. However, real-world applications present more complex and varied scenarios in which the quality and consistency of tracker signals cannot always be guaranteed. To address this, we have identified three primary scenarios in which our model and similar models may need to operate:
\begin{itemize}
    \item \textbf{Standard scenario:} This scenario represents ideal conditions where the model successfully receives all required tracker signals without any loss or degradation. This is the standard operational scenario for most existing methods.
    \item \textbf{Instantaneous data-loss scenario:} Real-world applications are prone to transient disruptions, such as network fluctuations, leading to the momentary loss of specific tracker signals. 
    To simulate these sudden and brief interruptions in signal transmission in this scenario, we make each tracker signal at any given time $t$ subject to a probability $p$ of being lost, as illustrated in Fig \ref{fig:simulation}.A. 

    \item \textbf{Prolonged data-loss scenario:} There are instances where certain tracker signals may be consistently unavailable over a long-lasting period\footnote{The term ``long-lasting period" refers to a period that is significantly longer than the frame interval, for example, 1 second.}. Such situations can arise from continuous occlusion of infrared markers or sustained invisibility of hand-held controller signals in an HMD kit. This scenario is characterized by a continuous loss of all tracker signals for a specific duration, as depicted in Fig \ref{fig:simulation}.B. It represents challenges posed by prolonged data unavailability.
\end{itemize}

The detailed configurations and implications of these scenarios will be further explored in Section \ref{sec:simulation} for the training stage and Section \ref{sec:protocols} for the testing stage, respectively.

\subsection{Overview}
ReliaAvatar adopts an autoregressive pipeline during both the training and testing stages. The process of ReliaAvatar's handling continuous tracker signal sequences can be represented by the following equations:
\begin{equation}
    \hat{y}^0,h_x^0,h_y^0=f_{\theta}(x^0)
\end{equation}
\begin{equation}
    \hat{y}^1,h_x^1,h_y^1=f_{\theta}(x^1,\hat{y}^{0},h_x^{0},h_y^{0})
\end{equation}
\begin{equation*}
    ...
\end{equation*}
\begin{equation}
    \hat{y}^{t},h_x^{t},h_y^{t}=f_{\theta}(x^{t},\hat{y}^{t-1},h_x^{t-1},h_y^{t-1})
\end{equation}
\begin{equation*}
    ...
\end{equation*}
During the training stage, given a tracker signal sequence ${\{x^t\}}_{t=0}^{L-1}$ as input, the training process is divided into $L$ sequential steps with each step building upon the previous one. $L$ is the length of the sequence.

As depicted in Figure \ref{fig:pipeline}, 
our model features two distinct yet interconnected pathways: the regression pathway (Regression Encoder$\rightarrow$Joint-relation Transformer$\rightarrow$Decoder) and the prediction pathway (Prediction Encoder$\rightarrow$Joint-relation Transformer$\rightarrow$Decoder). The regression pathway operates similarly to the traditional avatar animator. Its primary function is to regress full-body poses from sparse joint information. This process can be formulated as:

\begin{equation}
    \hat{y}^t,h_x^t=f^{reg}_{\theta}(x^t,h_x^{t-1})
\end{equation}

The prediction pathway, on the other hand, aligns more closely with motion prediction models. It utilizes known sequences of past motion to predict the current full-body motion. The formulation for this pathway is as follows:
\begin{equation}
    \hat{y}^t,h_y^t=f^{pred}_{\theta}(\hat{y}^{t-1},h_y^{t-1})
\end{equation}

Both pathways converge at the Joint-Relation Transformer. This convergence results in the complete model represented by Eq.~\eqref{eq:full_formulation}. The Joint-Relation Transformer and the decoder are shared modules utilized by both pathways, while each pathway has its unique components – the regression encoder for the regression pathway and the prediction encoder for the prediction pathway.

\subsection{Model Architecture}
{\flushleft\textbf{Regression Encoder and Prediction Encoder.}}
As shown in Figure \ref{fig:pipeline}, 
tracker signals are initially processed through a sparse fully connected layer (SFC) to extract initial features $E_0^t\in \mathbb{R}^{|\mathbb{J}|\times d}$, where $d$ is the embedding dimension of the model. The purpose of SFC is to merge multiple linear layers into a sparse linear layer, thereby reducing the number of loops in the forward process and improving the runtime efficiency. In the regression encoder, SFC is formed by merging four linear layers that handle orientation $x_r$, rotation velocity $\Delta{x}_r$, position $x_p$, and linear velocity $\Delta{x}_p$. This module can be formulated as:
\begin{equation}
    E_0 = \begin{bmatrix}
x_r \\ \Delta{x}_r \\ x_p \\ \Delta{x}_p
\end{bmatrix}
\begin{bmatrix} 
W_1^T & \textbf{0} &\textbf{0}&\textbf{0} \\ 
\textbf{0}& W_2^T &\textbf{0} &\textbf{0} \\
\textbf{0}&\textbf{0} &W_3^T &\textbf{0} \\
\textbf{0}& \textbf{0}&\textbf{0}&W_4^T 
\end{bmatrix}+b
\end{equation}
Then, the GRU module will extract the temporal features, as described in the following formula:
\begin{equation}
    E_x^t, h_x^t = GRU(E_0^t, h_x^{t-1})
\end{equation}
After the regression encoder, the features of all tracker signals $E_x^{t} \in \mathbb{R}^{|\mathbb{J}|\times d}$ are obtained. Subsequently, an upsampling module will be used to increase the dimension to $22\times d$, thus obtaining the decoding token sequence $\{t_x^i\}_{i=0}^{21}$ which will be input to the Joint-Relation Transformer. Each token represents a SMPL joint.

The prediction encoder is similar to the regression encoder. After passing through an SFC and a GRU module, $\hat{y}^{t-1}$ is mapped to a token $E_y^t\in \mathbb{R}^{d}$. Then a fully connected layer is applied to expand it to 22 joint tokens $\{t_y^i\}_{i=0}^{21}$.

{\flushleft\textbf{Joint-Relation Transformer.}}
In human kinematics, the motions of the entire body are not simply a combination of independent motions of multiple joints. In contrast, the actions performed by the human body are strongly correlated with the relationship between joints. Therefore, it is evident that equipping the model with the ability to model joint relationships can enhance the driving effectiveness. A Transformer encoder is used to model joint relationships, which has been proven to be effective in PETR~\cite{Shi_2022_CVPR}.

The input of the Transformer is a token sequence $\{t^i\}_{i=0}^{21}=$ Concat $(t_x,t_y) \in \mathbb{R}^{22\times (2d)}$ representing 22 SMPL joints, which is fused from the tokens generated by the previous two encoders.

{\flushleft\textbf{Decoder.}}
The decoder is responsible for decoding rotation $\hat{y}_{rot}^t$ and position $\hat{y}_{pos}^t$ directly from the output of the token sequence of the Transformer. The decoder is composed of two layers of MLP. There are three decoder variants:
\begin{itemize}
    \item \textbf{Shared Decoder.} Since the pelvic joint parameter represents the global orientation of the entire body, while the other joint parameters represent local rotations relative to their parent joints, a dedicated decoder is used for the pelvic joint, while the remaining joints share a common decoder.
    \item \textbf{Multi-FC Decoder.} Each joint has its own dedicated decoder. During the inference process, the decoding of each joint is done sequentially.
    \item \textbf{SFC Decoder.} Use an MLP composed of SFCs as a unique decoder. As a result, all joints can be decoded simultaneously in parallel.
\end{itemize}

We adopted SFC decoder due to its advantages of high accuracy and high efficiency. The comparison of the three decoders will be mentioned in Sec \ref{sec:ablation}.

\subsection{Simulation Training}\label{sec:simulation}
We simulate the three scenarios mentioned in Sec \label{sec:definition} during training ReliaAvatar. 

\begin{itemize}
    \item \textbf{Standard scenario.} In the standard scenario, all tracker signals can be fully and correctly transmitted to the model. So in this scenario, there is no need for any masking of the signals.
    \item \textbf{Instantaneous data-loss scenario.} In the instantaneous data-loss scenario, we use a dropout layer to randomly mask the signals of each tracker in each frame with a probability of 0.1.
    \item \textbf{Prolonged data-loss scenario.} In the prolonged data-loss scenario, we mask out the latter half of each input sequence $\{x_t\}_{t=0}^{L-1}$. That is to say, $\{x_t\}_{t=L/2}^{L-1}=$ \textbf{0}.
\end{itemize}
Each sequence has an equal probability ($1/3$) to undergo the above three simulation treatments.

\subsection{Loss Function}
We adopt orientation loss $\mathscr{L}_{ori}$, rotation loss $\mathscr{L}_{rot}$, SMPL position loss $\mathscr{L}_{pos}^{SMPL}$, decoded position loss $\mathscr{L}_{pos}^{dec}$, and velocity loss $\mathscr{L}_{vec}$ during the optimization process. All losses are computed using the L1 loss. Taking the rotation loss as an example, its loss function is as follows:
\begin{equation}
    \mathscr{L}_{rot}=\frac{\Sigma_{t=0}^{L-1}||\hat{y}_{rot}^t-y_{rot}^t||}{L}
\end{equation}
Here, $\hat{y}_{rot},y_{rot}$ represent the model output and corresponding ground truth, respectively. 
Other losses are computed in the same way.
The whole loss function can be formulated as:
\begin{equation}
    \begin{aligned}
    \mathscr{L}=&\lambda_{ori}\mathscr{L}_{ori} + \lambda_{rot}\mathscr{L}_{rot} + \lambda_{pos}^{SMPL}\mathscr{L}_{pos}^{SMPL} + \\
    &\lambda_{pos}^{dec}\mathscr{L}_{pos}^{dec} + 
    \lambda_{vec}\mathscr\mathscr{L}_{vec}
    \end{aligned}
\end{equation}
Here, $\lambda_{ori},\lambda_{rot},\lambda_{pos}^{SMPL},\lambda_{pos}^{dec},\lambda_{vec}$ are weights of corresponding losses.
The difference between $\mathscr{L}_{pos}^{SMPL}$ and $\mathscr{L}_{pos}^{dec}$ is that $\mathscr{L}_{pos}^{SMPL}$ is computed based on the position $SMPL(\hat{y}_{rot})$ inferred by SMPL and the ground truth $y_{pos}$, while $\mathscr{L}_{pos}^{dec}$ is computed based on the position $\hat{y}_{pos}$ decoded by the decoder and $y_{pos}$.
\section{Experiments}
\subsection{Implementation Details}
For all GRUs in the model, the number of layers is set to 1 and the dimension is set to 256. The joint-relation Transformer has 4 layers with a dimension of 512. The initial learning rate is set to $5\times 10^{-4}$, 
and is halved every 15000 iterations. The length of the input sequence $L$ is set to 32. $\{\lambda_{ori},\lambda_{rot},\lambda_{pos}^{SMPL},\lambda_{pos}^{dec},\lambda_{vec}\}=\{0.02,1,1,1,0.5\}$. The model is trained on two GeForce RTX 3090 GPUs for a total of 90000 iterations, with a batch size of 32 on each GPU. The training process costs approximately 32 hours to complete. 

\subsection{Evaluation Metrics}
We adopt the following evaluation metrics.
\vspace{2pt}
\begin{itemize}
    \item \textbf{MPJPE}: Mean Per Joint Position Error [cm].
    \item \textbf{MPJRE}: Mean Per Joint Rotational Error [degree].
    \item \textbf{MPJVE}: Mean Per Joint Velocity Error [cm/s].
    \item \textbf{fps}: Frames Per Second [frame].
\end{itemize}

\subsection{Evaluation Protocols}\label{sec:protocols}
{\flushleft\textbf{Standard scenario.}}
Following the experimental setup outlined in AvatarPoser~\cite{jiang2022avatarposer}, we partition the CMU~\cite{CmuMocap}, BMLrub~\cite{troje2002decomposing}, and HDM05~\cite{muller2007mocap} subsets into 90\% training data and 10\% testing data. For comparison, we evaluate the results using both three inputs (head and wrists) and four inputs (head, wrists, and pelvis). It is worth noting that, unless explicitly stated otherwise, all other results are based on the usage of three inputs.

{\flushleft\textbf{Instantaneous data-loss scenario.}}
AGRoL~\cite{du2023avatars} proposed a setting for instantaneous data-loss, which randomly masks out 10\% of the input sequence. However, in practical applications, an instantaneous data-loss does not necessarily mean the complete loss of all tracker signals. Instead, it might indicate the loss of information from a specific joint. Therefore, 
we propose a new setting where signals of each tracker signal have a certain probability $p$ of being lost at each moment. We evaluated each model five times at $p=0.1, 0.5$ and $0.9$, and took the average as the result.

{\flushleft\textbf{Prolonged data-loss scenario.}}
In the scenario of prolonged data-loss, data is continuously lost for a long-last period. For evaluation purposes, we set a protocol to mask out subsequent $M$ frames every 80 frames. We evaluated each model in $M=20, 40$ and $60$.

\subsection{Comparison to the State-of-the-art}
The evaluation results of each model in the standard scenario are extracted from the reports in their respective papers. For comparison in data-loss scenarios, we select several publicly available state-of-the-art methods (AvatarPoser~\cite{jiang2022avatarposer}, AGRoL~\cite{du2023avatars} and AvatarJLM~\cite{zheng2023realistic}). We conduct a fair comparison by retraining their publicly available code and evaluating them under the protocols for data-loss scenarios using the retrained parameters.
{\flushleft\textbf{Standard scenario.}} To verify the effectiveness of our model, we perform a fair comparison with the state-of-the-art methods in the standard scenario. 
The results, as shown in Table \ref{tab:three_input} and Table \ref{tab:four_input}, demonstrate that our model achieves state-of-the-art performance in three- and four-input conditions. This indicates that ReliaAvatar, despite being primarily designed to tackle data-loss scenarios, surpasses other models even in the standard scenario.

\begin{table}[h]\small
\centering
\resizebox{0.49\textwidth}{!}{
\begin{tabular}{c|ccc}
\toprule
 Method & MPJRE & MPJPE & MPJVE\\ 
\midrule 
FinalIK~\cite{finalIK}&16.77&18.09 &59.24\\
CoolMoves~\cite{ahuja2021coolmoves}&5.20 & 7.83&100.54\\
LoBSTr~\cite{yang2021lobstr}&10.69 &9.02 & 44.97\\
VAE-HMD~\cite{dittadi2021full} & 4.11 &6.83 & 37.99\\
AvatarPoser~\cite{jiang2022avatarposer} & 3.21& 4.18&29.40\\
EgoPoser~\cite{jiang2023egoposer} &- &4.14 &25.95\\
AGRoL~\cite{du2023avatars}  & 2.66 & 3.71& 18.59 \\
DAP~\cite{di2023dual} & 2.69 & 3.68 & 24.30\\
{AvatarJLM}~\cite{zheng2023realistic} & {2.90}& {3.35} & {20.79}\\
ReliaAvatar (Ours) & \textbf{2.53} &\textbf{3.18} &\textbf{18.30}  \\
\bottomrule
\end{tabular}
}
\caption{State-of-the-art comparison on AMASS in the standard scenario using three inputs ({head and wrists}). Best in {bold}.}
\label{tab:three_input}
\end{table}

\begin{table}[h]\small
\centering
\resizebox{0.49\textwidth}{!}{
\begin{tabular}{c|ccc}
\toprule
Method & MPJRE & MPJPE & MPJVE\\ 
\midrule 
FinalIK~\cite{finalIK} &12.39 &9.54 & 36.73\\
CoolMoves~\cite{ahuja2021coolmoves}& 4.58& 5.55&65.28\\
LoBSTr~\cite{yang2021lobstr}&8.09 &5.56 &30.12\\
VAE-HMD~\cite{dittadi2021full} & 3.12 &3.51 & 28.23\\
AvatarPoser~\cite{jiang2022avatarposer} &2.59 &2.61 &22.16 \\
{AvatarJLM}~\cite{zheng2023realistic}&{2.40} & {2.09}& {17.82} \\
ReliaAvatar (Ours)&\textbf{2.16} &\textbf{1.94} &\textbf{14.32}\\

\bottomrule
\end{tabular}
}
\caption{State-of-the-art comparison on AMASS in the standard scenario using four inputs ({head, wrists, and pelvis}). Best in {bold}.}
\label{tab:four_input}
\end{table}

{\flushleft\textbf{Instantaneous data-loss scenario.}}
We compare ReliaAvatar with other models under three conditions: $p=0.1, p=0.5$, and $p=0.9$. As shown in Table \ref{tab:short-term}, ReliaAvatar demonstrates its superior robustness compared to other models in the instantaneous data-loss scenario. Under the conditions of $p=0.1$ and $0.5$, ReliaAvatar is hardly affected. Even when the signals of each tracker have only a 10\% probability of being received, ReliaAvatar can still operate normally.
\begin{table}\small
\centering
\begin{tabular}{c|c|ccc}
\toprule
$p$ & Method & MPJRE & MPJPE & MPJVE \\
\midrule
\multirow{4}{*}{0.1}
& AvatarPoser &9.01 &20.02 &1532.62 \\
& AGRoL &\underline{6.59}& \underline{12.31}& \underline{101.71} \\
& {AvatarJLM} &6.65 &13.40 & 887.24\\
& ReliaAvatar(Ours) &\textbf{2.66} &\textbf{3.32} & \textbf{20.44} \\
\midrule
\multirow{4}{*}{0.5}
& AvatarPoser &15.47 & 53.07& 3651.32\\
& AGRoL &\underline{9.96} & \underline{20.04}&\underline{114.12} \\
& {AvatarJLM} & 16.86& 43.02& 2512.13\\
& ReliaAvatar(Ours) &\textbf{2.77} &\textbf{3.51} &\textbf{26.51} \\
\midrule
\multirow{4}{*}{0.9}
& AvatarPoser &17.98 &63.53 & 2582.13\\
& AGRoL &\underline{13.08} &\underline{27.59} &\underline{117.26} \\
& {AvatarJLM} & 25.16& 65.50& 1154.16 \\
& ReliaAvatar(Ours) &\textbf{5.12} &\textbf{7.86} &\textbf{45.00} \\
\bottomrule
\end{tabular}
\caption{
Comparison in instantaneous data-loss scenario. Best in {bold} and second best {underlined}.}
\vspace{-0.05in}
\label{tab:short-term}
\end{table}

{\flushleft\textbf{Prolonged data-loss scenario.}}
We compare our model with the previous state-of-the-art models in the prolonged data-loss scenario under three conditions: $M=20, 40$, and $60$. Table \ref{tab:long-term} indicates that existing models are unable to handle scenarios where there is a continuous loss of more than 40 frames of signals. In contrast, ReliaAvatar exhibits superior robustness compared to other methods, as it can function nearly normally even with a continuous loss of 20 frames. It is capable of operating even when the signal is continuously lost for 60 frames.

ReliaAvatar's robustness to low-quality signals stems from three aspects: 1) Our model abandons windowed inputs. Models that solely rely on a window of tracker signals as input (e.g., AvatarPoser with a window size of 40) exhibit obvious malfunctions when facing prolonged data-loss since the signals within the window are all set to zero. 2) Our model integrates motion prediction into the avatar animator. When the regression pathway fails to function properly, the prediction pathway can still predict the current full-body motion based on the historical trajectory states. 3) The autoregressive training paradigm allows for simulating abnormal scenarios that may occur in applications.
\begin{table}\small
\centering
\begin{tabular}{c|c|ccc}
\toprule
$M$ & Method & MPJRE & MPJPE & MPJVE \\
\midrule
\multirow{4}{*}{20}
& AvatarPoser &7.43 &14.46 &102.18 \\
& AGRoL &\underline{6.59}& \underline{12.31}& \underline{101.71} \\
& {AvatarJLM} &8.96 &18.71 & 133.92\\
& ReliaAvatar (Ours) &\textbf{2.84} &\textbf{3.67} & \textbf{24.69} \\
\midrule
\multirow{4}{*}{40}
& AvatarPoser &11.52 & 23.57& 123.33\\
& AGRoL &\underline{9.96} & \underline{20.24}&\underline{114.12} \\
& {AvatarJLM} & 14.52&34.39 & 154.17\\
& ReliaAvatar (Ours) &\textbf{3.48} &\textbf{5.17} &\textbf{34.01} \\
\midrule
\multirow{4}{*}{60}
& AvatarPoser &15.05 &31.36 & 118.73\\
& AGRoL &\underline{13.08} &\underline{27.59} &\underline{117.26} \\
& {AvatarJLM} & 20.14& 49.95& 181.36 \\
& ReliaAvatar (Ours) &\textbf{4.70} &\textbf{7.69} &\textbf{45.02} \\
\bottomrule
\end{tabular}
\caption{Comparison in prolonged data-loss scenario. Best in {bold} and second best {underlined}.
}
\vspace{-0.05in}
\label{tab:long-term}
\end{table}

\subsection{Ablation Studies}
\label{sec:ablation}

To illustrate the roles of various designs in our model, we conduct the following ablation experiments.

{\flushleft\textbf{Input Signals.}} As shown in Table \ref{tab:input_signals}, we validate the performance gain of incorporating historical trajectory states into the input. Here, $X$ represents the input used by most current methods, i.e. $X=\{x_r,\Delta{x}_r,x_p,\Delta{x}_p\}$. 

If the historical trajectory states are not used as input, the regression pathway remains inactive. The experimental results demonstrate that incorporating $y_{rot},\Delta{y}_{rot},y_{pos},\Delta{y}_{pos}$ into the input leads to performance gains. Due to the continuity of motion, the current full-body is naturally influenced by and aligned with the previous trajectories. So this improvement can be attributed to the previous trajectories providing cues for the generation of the current full-body motion.

\begin{table}\small
\centering
\resizebox{0.49\textwidth}{!}{
\begin{tabular}{c|ccc}
\toprule
         Input signals& MPJRE& MPJPE & MPJVE \\
         \midrule
         $\{X\}$&2.69 & 3.49& 23.49 \\
         $\{X,y_{rot}\}$&2.54 &3.25 &19.44 \\
         $\{X,y_{pos}\}$&2.58 &3.28& 18.39\\
         $\{X,y_{rot},y_{pos}\}$&2.53 &3.25 &19.07\\

         $\{X,y_{rot},y_{pos},\Delta{y}_{rot},\Delta{y}_{pos}\}$&\textbf{2.53} & \textbf{3.18}&\textbf{18.30}\\
\bottomrule 
\end{tabular}
}
\caption{Ablation experiments of the input signals. Best in {bold}.}
\label{tab:input_signals}
\end{table}

{\flushleft\textbf{Decoder.}}
As shown in Table \ref{tab:decoder}, we compare the performance of different decoder designs in our model. The model using a shared decoder has larger errors but shorter runtime while the model using multi-FC decoders has smaller errors but suffers from poor real-time performance. The model using a Sparse-FC decoder has the advantages of both low error and fast running speed.

\begin{table}\small
\centering
\begin{tabular}{c|cccc}
\toprule
         Decoder& MPJRE& MPJPE & MPJVE & fps \\
         \midrule
         Shared &2.64 & 3.38& 19.51& \textbf{112.22}\\
         Multi-FC & \textbf{2.51}&\underline{3.21} & 18.64&46.73\\
         Sparse-FC &\underline{2.53} & \textbf{3.18}&\textbf{18.30}&\underline{109.65}\\
\bottomrule 
\end{tabular}
\caption{Ablation experiments of the design of decoder. Best in {bold} and second best {underlined}.}
\label{tab:decoder}
\end{table}

{\flushleft\textbf{Fusion Method.}}
The methods to integrate features from two pathways are worth exploring. We have explored three different methods for feature integration: addition, concatenation along the feature dimension, and concatenation along the joint dimension.
As shown in Table \ref{tab:fusion}, the fusion method of concatenating features along the embedding dimensions achieves the best performance.
\begin{table}[]\small
\centering
\begin{tabular}{c|ccc}
\toprule
         method& MPJRE& MPJPE & MPJVE  \\
         \midrule
         add &2.66 &3.31 & 19.70\\
         concat (joint-dim) &2.66 & 3.31&19.83 \\
         concat (feat-dim) &\textbf{2.53} &\textbf{3.18} &\textbf{18.90} \\
\bottomrule 
\end{tabular}
\caption{{Ablation experiments of the fusion method.} Best in {bold}.}
\label{tab:fusion}
\end{table}

{\flushleft\textbf{Simulation Training.}}
An important distinction of Relia\-Avatar compared to other methods is its ability to simulate various scenarios during the training process. This is attributed to the model architecture with joint motion prediction and the autoregressive training paradigm. The results presented in Table \ref{tab:training} demonstrate that simulation training for data-loss scenarios does not adversely affect performance in the standard scenario.
\begin{table}\small
\centering
\begin{tabular}{c|ccc}
\toprule
         method& Standard.& Instantaneous & Prolonged   \\
         \midrule
         no mask &\textbf{3.17} & 51.41& 24.30\\
         + random mask &3.18 & \underline{3.92} &\underline{12.26} \\
         + prolonged mask &\underline{3.18}&\textbf{3.51}&\textbf{5.17} \\
\bottomrule 
\end{tabular}
\caption{Ablation experiments of simulation training. Best in {bold} and second best {underlined}.}
\label{tab:training}
\end{table}

\subsection{Analysis}
{\flushleft\textbf{Ability of motion prediction}.}
An important capability that we expect ReliaAvatar to achieve is the ability to predict full-body movements of the current frame using historical trajectory states when tracker signals are lost. 
This enables us to drive avatars smoothly even in the absence of real-time tracker information.
We calculate MPJPE for the 40 frames after losing real-time tracker signals. The results presented in Table \ref{tab:prediction} demonstrate that ReliaAvatar is capable of predicting relatively accurate poses within several tens of frames after the disappearance of tracker signals. 
\begin{table}\small
\centering
\begin{tabular}{c|cccccccccc}
\toprule
         Frame& \#1&\#3&\#7&\#10&\#20&\#30&\#40  \\
         \midrule
         MPJPE & 3.41 & 3.51&3.91 & 4.34& 6.34& 8.94 &11.62\\
\bottomrule 
\end{tabular}
\caption{The statistical errors after the interruption of tracker signals. \#n represents the n-th frame after the interruption.}
\label{tab:prediction}
\end{table}

\begin{figure}[h]
    \centering
    \begin{tikzpicture}[/pgfplots/width=0.95\linewidth,
    /pgfplots/height=0.6\linewidth,
    font=\small]
    \begin{axis}
    [ymin=3,ymax=4.6,xmin=0,xmax=45,
    xtick={0,5,10,15,20,25,30,35,40},
    xticklabels={0,,10,,20,,30,,40},
    ytick={3,3.4,3.8,4.2,4.6},
    yticklabels={3,3.4,3.8,4.2,4.6},
    xlabel={Running Time (ms)},
    ylabel={MPJPE (cm)},
    y label style={yshift=-12pt},
    legend cell align={left},
    legend pos=north west,
    grid=both,
    grid style=dotted,
    major grid style={white!20!black},
    minor grid style={white!70!black},
    nodes near coords,
    point meta=explicit symbolic,
    every node near coord/.append style={font=\footnotesize,anchor=west}
    ]
    \addplot[
      mark=*,
      only marks,
    ] coordinates {
      (9.12, 3.18) [ReliaAvatar (Ours)]
    };
    \addplot[
      mark=*,
      only marks,
      mark options={color=blue}, 
    ] coordinates {
      (9.12, 3.18) 
    };
    \addplot[
      mark=*,
      only marks,
    ] coordinates {
      (12.43, 4.18) [AvatarPoser]
    };
    \addplot[
      mark=*,
      only marks,
      mark options={color=red}, 
    ] coordinates {
      (12.43, 4.18) 
    };
    \addplot[
      mark=*,
      only marks,
    ] coordinates {
      (32.79, 3.71) [AGRoL]
    };
    \addplot[
      mark=*,
      only marks,
      mark options={color=green}, 
    ] coordinates {
      (32.79, 3.71) 
    };
    \addplot[
      mark=*,
      only marks,
    ] coordinates {
      (32.69, 3.35) [AvatarJLM]
    };
    \addplot[
      mark=*,
      only marks,
      mark options={color=orange}, 
    ] coordinates {
      (32.69, 3.35) 
    };
    \end{axis}
    \end{tikzpicture}
    \caption{Inference time comparison.}
    \label{fig:inference}
\end{figure}
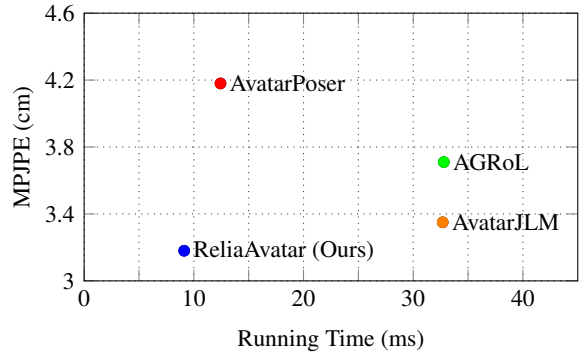

{\flushleft\textbf{Inference Time.}} To ensure a fair comparison of the operational efficiency of different models, we conduct speed tests on state-of-the-art models under identical operating modes and hardware conditions. For each model, we construct a zero input and run it 10,000 times to obtain the average time as its inference time. All tests are conducted on a single GeForce RTX 3090 GPU. As shown in Figure \ref{fig:inference}, ReliaAvatar surpasses other models in both operational efficiency and accuracy.

\section{Conclusion}
We conducted a comprehensive exploration of the full-body avatar pose estimation problem under low-quality signal scenarios, which had not been systematically investigated before.
We summarize three scenarios that may be encountered in practical applications: standard scenario, instantaneous data-loss scenario, and prolonged data-loss scenario.
To address these challenges, we proposed \textit{ReliaAvatar}, a real-time, robust, autoregressive avatar animator. We consider the full-body avatar pose estimation problem as a combination of joint upsampling and motion prediction. Therefore, ReliaAvatar possesses an upsampling pathway and a prediction pathway. Furthermore, we incorporate simulation training for data-loss scenarios on top of autoregressive training.
Experimental results demonstrate that ReliaAvatar not only outperforms other methods in the data-loss scenarios but also achieves state-of-the-art performance in the standard scenario. In addition to its outstanding accuracy and robustness, ReliaAvatar also exhibits superior running efficiency compared to other methods.
These advantages reduce the hardware requirements for animating a full-body avatar. As a result, more affordable and convenient devices can be used to drive avatars, which contributes to the wider adoption of related technologies.

\newpage
\section*{Acknowledgements}
This work was supported by National Key R\&D Program of China under Grant No. 2021ZD0110400, the Fundamental Research Funds for the Central Universities No. xxj032023020, and sponsored by the CAAI-MindSpore Open Fund, developed on OpenI Community.

\bibliographystyle{named}
\bibliography{ijcai24}

\end{document}